\documentclass[3p,times,procedia]{elsarticle}
\usepackage{ecrc}
\usepackage{amsmath}
\usepackage{caption}
\usepackage{subcaption}
\usepackage{svg}
\usepackage{tikz}

\volume{00}

\firstpage{1}

\journalname{Transportation Research Procedia}

\runauth{D. Dekker et al.}

\jid{trpro}

\usepackage{amssymb}

\biboptions{authoryear}

\usepackage[figuresright]{rotating}

\usepackage[bookmarks=false]{hyperref}
    \hypersetup{colorlinks,
      linkcolor=blue,
      citecolor=blue,
      urlcolor=blue}

\begin{document}
\begin{frontmatter}



\dochead{26th Euro Working Group on Transportation Meeting (EWGT 2024)}%

\title{A new method for inserting train paths into a timetable}
\title{A New Method for Inserting Train Paths into a Timetable}


\author[a]{David Dekker\corref{cor1}}
\author[a]{Carl Henrik Häll}
\author[a]{Anders Peterson}
\author[a]{Christiane Schmidt}

\address[a]{Communications and Transport Systems, ITN, Linköping University, Sweden}

\begin{abstract}
A seemingly simple, yet widely applicable subroutine in automated train scheduling is the insertion of a new train path to a timetable in a railway network.
We believe it to be the first step towards a new train-rerouting framework in case of large disturbances or maintenance works.
Other applications include handling ad-hoc requests and modifying train paths upon request from railway undertakings.
We propose a fast and scalable path-insertion algorithm based on dynamic programming that is able to output multiple suitable paths.
Our algorithm uses macroscopic data and can run on railway networks with any number of tracks.
We apply the algorithm on the line from Göteborg Sävenäs to the Norwegian border at Kornsjö.
For a time window of seven hours, we obtain eight suitable paths for a freight train within $0.3$ seconds after preprocessing.
\end{abstract}

\cortext[cor1]{Corresponding author. Tel.:  +31 6 37 16 90 89 .}
\end{frontmatter}

\email{david.dekker@liu.se}


\section{Introduction}
\label{intro}
Optimization techniques have been widely applied to solve problems in railway timetabling, as explained in, for example,~\cite{opthandbook}.
One of the most basic problems is the addition of one new train path to a timetable.
The problem has many applications for infrastructure managers, like processing ad-hoc requests, altering paths when the train weight or locomotives change, and altering train paths when an early departure is requested.
Besides, we believe that an algorithm solving the path-insertion problem can be used as a subroutine in larger train-rerouting problems when multiple trains have to be rerouted.

Multiple approaches already exist for solving the path-insertion problem.
\cite{MIP} explore Mixed Integer Programs; \cite{ch} and \cite{kristian} consider shortest path algorithms on time-extended graphs and \cite{robustheuristics} apply ruin-and-recreate heuristics.
However, to the best of our knowledge, no form of automation is currently used in practice, despite the available solutions.

There can be several reasons for this difference.
One problem is the varying level of detail.
The infrastructure might be simplified too much in order to obtain usable results in practice, or trains might be required to always run at full speed.
We also need the algorithm to be fast, keeping in mind our intended use as a subroutine for larger rerouting problems.
Existing solutions using time-extended graphs often keep the timescale of the solution discrete, slowing down the algorithm significantly when the timetable is created at the level of seconds.
Lastly, the aforementioned approaches only output one (optimal) solution.
Obtaining multiple relevant options is useful in practice, which we discuss in Section~\ref{sec:results}.
In this paper, we present a new approach that achieves these benefits based on dynamic programming.
Our objective 
is to minimize travel time, and we assume that robustness measures are incorporated in the headway constraints.
More precisely, we aim to find all non-dominated paths, where a dominating path over a path with departure time $t_1$ and arrival time $t_2$ would have a departure $t_1' \ge t_1$ and arrival $t_2' \le t_2$, with at least one of the inequalities being strict.

\section{Terminology}
We use $\mathcal S$ to represent the set of all \emph{stations} under consideration, which are all relevant infrastructure points that are included in train paths.
The set $\mathcal L$ contains all (directed) \emph{segments}, the connections between stations.
Both stations and segments can have any number of tracks.
The function $f \colon \mathcal S \cup \mathcal L \rightarrow 2^{\mathbb N}$ returns the tracks at a station or segment.
For simplicity, we assume these tracks to be parallel, but the algorithm can be altered slightly to handle more complex station lay-outs.

We let $\mathcal A$ denote the set of trains. 
Each train $a \in \mathcal A$ operates on a set of stations $\mathcal S_a \subseteq \mathcal S$ and segments $\mathcal L_a \subseteq \mathcal L$.
We assume that each train $a \in \mathcal A$ visits each station at most once, which allows a simpler description of the algorithm.
To model route conflicts in station gridirons between stations and its entry signals, we introduce the notion of \emph{transitions}, denoted by $\mathcal T$.
A transition goes either from a station track to a segment track or vice versa.
We assume that we know which pairs of transitions conflict each other.
Meeting stations at single-track lines where no simultaneous arrivals are allowed, can also be modeled by conflicting transitions.

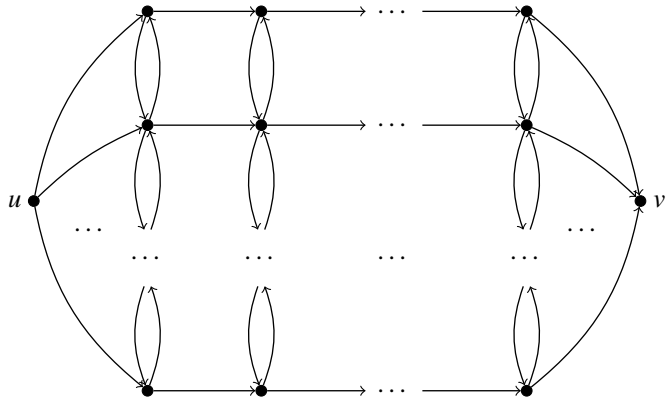
\begin{figure}[ht!]
\centering
\begin{tikzpicture}[circ/.style={circle,draw,inner sep=0pt,minimum size=4pt,fill=black}]
    \path[nodes={circ}]
    (0,0) node(u){}
    (1.5,2.5) node(a1){}
    (1.5,1) node(a2){}
    (1.5,-2.5) node(a5){}
    (3,2.5) node(b1){}
    (3,1) node(b2){}
    (3,-2.5) node(b5){}
    (6.5,2.5) node(e1){}
    (6.5,1) node(e2){}
    (6.5,-2.5) node(e5){}
    (8,0) node(v){}
    ;
    \node[] at (1.5,-0.5) (a3){};
    \node[] at (1.5,-1) (a4){};
    \node[] at (3,-0.5) (b3){};
    \node[] at (3,-1) (b4){};
    \node[] at (4.5,2.5) (c1){};
    \node[] at (4.5,1) (c2){};
    \node[] at (4.5,-0.5) (c3){};
    \node[] at (4.5,-1) (c4){};
    \node[] at (4.5,-2.5) (c5){};
    \node[] at (5,2.5) (d1){};
    \node[] at (5,1) (d2){};
    \node[] at (5,-0.5) (d3){};
    \node[] at (5,-1) (d4){};
    \node[] at (5,-2.5) (d5){};
    \node[] at (6.5,-0.5) (e3){};
    \node[] at (6.5,-1) (e4){};
    \node[] at (-0.25,0) {$u$};
    \node[] at (8.25,0) {$v$};
    \node[] at (0.75,-0.375) {$\dots$};
    \node[] at (1.5,-0.75) {$\dots$};
    \node[] at (3,-0.75) {$\dots$};
    \node[] at (4.75,-0.75) {$\dots$};
    \node[] at (4.75,2.5) {$\dots$};
    \node[] at (4.75,1) {$\dots$};
    \node[] at (4.75,-2.5) {$\dots$};
    \node[] at (6.5,-0.75) {$\dots$};
    \node[] at (7.25,-0.375) {$\dots$};

    \draw[line width=0.5pt,nodes={auto}]
    (u) edge[->,bend left=20] (a1)
    (u) edge[->,bend left=10] (a2)
    (u) edge[->,bend right=20] (a5)
    (a1) edge[bend right=20,->] (a2)
    (a2) edge[bend right=20,->] (a3)
    (a4) edge[bend right=20,->] (a5)
    (a5) edge[bend right=20,->] (a4)
    (a3) edge[bend right=20,->] (a2)
    (a2) edge[bend right=20,->] (a1)
    (a1) edge[->] (b1)
    (a2) edge[->] (b2)
    (a5) edge[->] (b5)
    (b1) edge[bend right=20,->] (b2)
    (b2) edge[bend right=20,->] (b3)
    (b4) edge[bend right=20,->] (b5)
    (b5) edge[bend right=20,->] (b4)
    (b3) edge[bend right=20,->] (b2)
    (b2) edge[bend right=20,->] (b1)
    (b1) edge[->] (c1)
    (b2) edge[->] (c2)
    (b5) edge[->] (c5)
    (d1) edge[->] (e1)
    (d2) edge[->] (e2)
    (d5) edge[->] (e5)
    (e1) edge[bend right=20,->] (e2)
    (e2) edge[bend right=20,->] (e3)
    (e4) edge[bend right=20,->] (e5)
    (e5) edge[bend right=20,->] (e4)
    (e3) edge[bend right=20,->] (e2)
    (e2) edge[bend right=20,->] (e1)
    (e1) edge[->,bend left=20] (v)
    (e2) edge[->,bend left=10] (v)
    (e5) edge[->,bend right=20] (v)
    ;
  \end{tikzpicture}
\caption{A graph class where there exists an arc ordering that is consistent with all $u$-$v$ paths. Such an ordering can, for example, be obtained by iteratively adding all arcs to the next layer, all downward arcs in its subsequent layer and all upward arcs in that layer. }\label{fig:nice-consistency}
\end{figure}

\begin{figure}[ht!]
\centering
\begin{subfigure}{.45\textwidth}
  \centering
  \begin{tikzpicture}[circ/.style={circle,draw,inner sep=0pt,minimum size=4pt,fill=black}]
    \path[nodes={circ}]
    (0,0) node(u){}
    (2,0) node(a1){}
    (4,0) node(a2){}
    (6,0) node(v){}
    (2,2) node(a3){}
    (4,2) node(a4){}
    (2,-2) node(a5){}
    (4,-2) node(a6){};
    \node[] at (-0.25,0) {$u$};
    \node[] at (6.25,0) {$v$};

    \draw[line width=0.5pt,nodes={auto}]
     (u) edge[->] (a1)
     (u) edge[bend left=25,->] (a3)
     (u) edge[bend right=25,->] (a5)
     (a3) edge[bend right=20,auto=right,->] (a1)
     (a1) edge[bend right=20,auto=right,->] (a5)
     (a5) edge[bend right=20,auto=right,->] (a1)
     (a1) edge[bend right=20,auto=right,->] (a3)
    (a3) edge[->] node{(1)} (a4)
    (a5) edge[->] node{(2)}(a6)
    (a2) edge[bend right=20,auto=right,->] (a1)
     (a1) edge[bend right=20,auto=right,->] (a2)
    (a4) edge[bend right=20,auto=right,->] (a2)
     (a2) edge[bend right=20,auto=right,->] (a6)
     (a6) edge[bend right=20,auto=right,->] (a2)
     (a2) edge[bend right=20,auto=right,->] (a4)
    (a2) edge[->] (v)
    (a4) edge[bend left=25,->] (v)
    (a6) edge[bend right=25,->] (v);
  \end{tikzpicture}
  \caption{A graph where no arc ordering is consistent with all $u$-$v$ paths.}
  \label{fig:init-a}
\end{subfigure}%
\begin{subfigure}{.45\textwidth}
  \centering
  \begin{tikzpicture}[circ/.style={circle,draw,inner sep=0pt,minimum size=4pt,fill=black}]
    \path[nodes={circ}]
    (0,0) node(u){}
    (2,0) node(a1){}
    (4,0) node(a2){}
    (6,0) node(v){}
    (2,2) node(a3){}
    (4,2) node(a4){}
    (2,-2) node(a5){}
    (4,-2) node(a6){};
    \node[] at (-0.25,0) {$u$};
    \node[] at (6.25,0) {$v$};

    \draw[line width=0.5pt,nodes={auto}]
     (u) edge[bend right=25,->] (a5)
     (a1) edge[->] (a3)
    (a3) edge[->] node[black]{(1)} (a4)
    (a5) edge[->] node[black]{(2)} (a6)
     (a2) edge[->] (a1)
     (a6) edge[->] (a2)
    (a4) edge[bend left=25,->] (v);
  \end{tikzpicture}
  \caption{Paths that cause the nonexistence of such an arc ordering.}\label{fig:inconsistent-path}
  \label{fig:init-b}
\end{subfigure}
\caption{By adding one arc to a graph from the graph class in Figure~\ref{fig:nice-consistency}, no arc ordering is consistent with all $u$-$v$ paths. Consider $u$-$v$ path and its mirrored counterpart in~\ref{fig:init-b}. The displayed path requires arc (2) before arc (1) in the ordering, while the mirrored path requires (1) before (2).}\label{fig:no-consistency}
\end{figure}
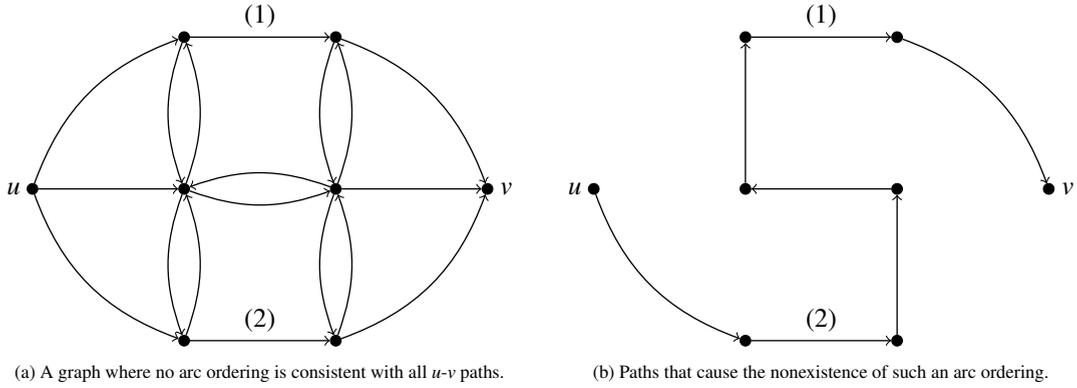

Lastly, we introduce the notion of \emph{arc orderings} that are \emph{consistent} with a $u$-$v$ path in a directed graph $D = (\mathcal S, \mathcal L)$ with vertices $u, v \in \mathcal S$.
Observe that this is not a train path, but a path in the graph-theoretical sense, i.e., a sequence of distinct vertices.
An \emph{arc ordering} $\sigma$ of a directed graph $D$ is a permutation of all arcs in $\mathcal L$.
It is \emph{consistent} with a $u$-$v$ path in $D$ if the vertices in that path occur in the same order in $\sigma$.
Some graph classes have the property that some arc ordering is consistent with all $u$-$v$ paths  for a pair of vertices $u$ and $v$.
Directed acyclic graphs have that property for any pair of vertices, and Figure~\ref{fig:nice-consistency} illustrates another graph class with that property.
This example relies on having only forward arcs between the vertex layers: Figure~\ref{fig:no-consistency} displays a graph with one additional arc, causing no arc ordering to be consistent with all $u$-$v$ paths.
However, in practice, paths as displayed in Figure~\ref{fig:inconsistent-path} are rarely a reasonable alternative to the direct paths that avoid this added arc. 
In our algorithm, we only consider $u$-$v$ paths that are consistent with a given arc ordering, where $u$ is the origin and $v$ the destination.

\section{Data, Input Parameters and Modeling Choices}\label{sec:data}
We use the macroscopic Swedish timetable from 2021, which was kindly provided by Trafikverket.
For each train $a \in \mathcal A$ at each station $s \in \mathcal S_a$, the timetable contains the arrival time, departure time and planned track number.
As specified in the introduction, we aim to find train paths that minimize the travel time.
Mathematically, we are interested in one path for each pair of departure time $t_1$ and arrival time $t_2$ for which there does not exist a faster path with departure $t_1' \ge t_1$ and arrival $t_2' \le t_2$.
We work on a macroscopic level, but with additional conflict detection around stations to account for most microscopic conflicts as well.
We assume that for a pair of routes from a station track to a segment track, we know whether these routes are in conflict.
In practice, microscopic data about these routes might be required to determine the existence of these conflicts in larger stations.
We assume that the running time and headway parameters include robustness margins and that these are fixed.
Having sufficient headway does not have to be a symmetric property: on a segment, a high speed train typically needs a larger headway in front of it than the proceeding train requires behind it.
We introduce the following parameters for the margins and minimum running times, where $x$ denotes the train to be inserted.
They are used to determine when a segment can be used by train $x$, and are only defined if both trains use the indexed track or segment.
\begin{itemize}
    \item $\beta_{a, x, \ell}$ is the minimum headway between trains $a \in \mathcal A$ and $x$ on segment $\ell \in \mathcal L$.
    Analogously, we define $\beta_{x, a, \ell}$. 
    \item $\gamma_{a, x, s, j}$ is the minimum time difference between departure of train $a \in \mathcal A$ and arrival of train $x$ on track $j \in f(s)$ of station $s \in \mathcal S$.
    Analogously, we define $\gamma_{x, a, s, j}$.
    For readability, we assume that the parameters are independent of whether the trains stop or not.
    \item $\delta_{a, x, s, j, \ell, k}$ is the minimum time difference between train $a \in \mathcal A$ and the departure of train $x$ from station track $j \in f(s)$ towards segment track $k \in f(\ell)$.
    If train $a$ uses the infrastructure between tracks $j$ and $k$ upon departure, the time difference relates to the departure time of train $a$ at station $s$, and arrival otherwise.
    When $x$ arrives at the station track from the segment track instead of departing, we slightly abuse notation and use $\delta_{a, x, \ell, k, s, j}$ to denote the time difference.
    These parameters are only defined if the route of train $a$ conflicts with a route from track $j$ to $k$.
    For simplicity, we assume that at most one of those two actions yields a conflict at the transition, making the provided indices sufficient to derive the minimum time differences.
    In practice, that is not always the case: for example, when two trains on a single-track line meet on a station that does not allow simultaneous arrivals, both the departure and the arrival of train $a$ causes a conflict with the arrival of train $x$.
    \item $d_{x, s_1, s_2}^\text{RR}$ is the minimum running time for train $x$ from station $s_1 \in \mathcal S$ to an adjacent station $s_2 \in \mathcal S$ if the train runs through both stations.
    When the train stops at the first station, last station, or both stations, we use $d_{x, s_1, s_2}^\text{SR}$, $d_{x, s_1, s_2}^\text{RS}$, and $d_{x, s_1, s_2}^\text{SS}$, respectively.
\end{itemize}
The running time data was provided by Trafikverket.
Regarding headways, they currently require a minimum of one minute between (microscopic) \emph{block occupations}.
We use three minutes for the minimum headways, as was also used by~\cite{3min}.
The $\delta$-parameter deviates when the first train arrives at the station, while the second train departs.
In this case, we are in the situation of comparing block occupations, so we use one minute instead.

\section{The Path Insertion Algorithm}

Our algorithm consists of three phases.
First, we compute when each station and segment can be used.
The dynamic program then determines how these free spaces can be traversed by train paths.
Lastly, we scan the tables filled by the dynamic program to construct the paths.

\subsection{Algorithm Initialization}
Let $D = (\mathcal S, \mathcal L)$ be the directed graph representing the railway network, i.e., with an arc in both directions whenever two stations are adjacent.
The reason for using this directed representation, is because we require an arc ordering of $\mathcal L$ as an input for our algorithm.
Let $u, v \in \mathcal S$ be the origin and destination of the train under consideration.
The algorithm subsequently considers all train paths for which the $u$-$v$ path through the network is consistent with the arc ordering.
An arc ordering that is consistent with the most relevant $u$-$v$ paths can be constructed greedily using $k$-shortest path algorithms like Yen's algorithm~\cite{yen}.
If an arc is not in any of the considered shortest paths, it can be removed from the network to speed up the computations.
We restrict ourselves to a given time interval $[T_{\min}, T_{\max}]$.
For some stations, there might be additional requirements, like a minimum stopping time or bounds on the arrival or departure.
These constraints can be enforced in the algorithm, but we leave the details to the reader.

For all stations, segments and transitions to be traversed, we first determine the \emph{free intervals} that can be used by the train without causing conflicts.
To determine these intervals, we use the parameters that were defined in Section~\ref{sec:data}.
We let $F$ denote the function that obtains these intervals.
The exact type of output depends on the infrastructure element under consideration.
We distinguish the following variations, where $x$ again denotes the train to be inserted.
\begin{itemize}
    \item For a station $s \in \mathcal S$ and track $j \in f(s)$, the free intervals $F_{\mathcal S}(s, j)$ are all time intervals between $T_{\min}$ and $T_{\max}$ where the inserted train would have sufficient headway margins to any other train at this track, and vice versa.
    Let $a_1, a_2 \in \mathcal A$ be two consecutive trains on track $j \in f(s)$, where $a_1$ leaves at time $t_1$ and $a_2$ enters at time $t_2$.
    Then $[t_1 + \gamma_{a_1, x, s, j}, t_2 - \gamma_{x, a_2, s, j}]$, if nonempty, would be an interval in $F_{\mathcal S}(s, j)$.
    Recall that we assume that the intervals do not depend on whether train $x$ stops in station $s$.
    \item For a transition from track $j$ at station $s$ to track $k$ at segment $\ell$, the free intervals are denoted with $F_{\mathcal T}(s, j, \ell, k)$.
    With these intervals, we also account for conflicts with other transitions.
    The intervals contain all departure times for which the train can leave the station without causing conflicts.
    With slight abuse of notation again, we would use $F_{\mathcal T}(\ell, k, s, j)$ if train $x$ was entering the station from the segment.
    Let $a_1, a_2 \in \mathcal A$ be two consecutive trains whose routes conflict on the route from track $j$ in station $s$ to $k$ at segment $\ell$.
    Their arrival or departure times, corresponding to these conflicts, are $t_1$ and $t_2$.
    Then $[t_1 + \delta_{a_1, x, s, j, \ell, k}, t_2 - \delta_{x, a_2, s, j, \ell, k}]$, if nonempty, would be an interval in $F_{\mathcal T}(s, j, \ell, k)$.
    \item For segments, the intervals are slightly more complicated.
    In particular, for double-track segments, we need pairs of intervals: one interval for when the train can be at the start of the segment, and a corresponding one at the end.
    We assume that trains move at constant speed at a segment.
    For some consistency, we model single-track segments similarly, but observe that both intervals are identical when the segment has only one block.
    To obtain expressions for the intervals, we let $a_1, a_2 \in \mathcal A$ be two consecutive trains on the segment.
    Train $a_1$ enters the segment at time $t_1$ and leaves at $t_1'$, while train $a_2$ enters at $t_2$ and leaves at $t_2'$.
    Then $([t_1 + \beta_{a, x, \ell}, t_2 - \beta_{x, b, \ell}], [t_1' + \beta_{a, x, \ell}, t_2' - \beta_{x, b, \ell}])$ would be a pair of intervals in $F_{\mathcal L}(\ell, k)$, if both are nonempty.
    Otherwise, we discard the interval pair, as the train cannot enter or leave the segment between trains $a_1$ and $a_2$.

\end{itemize}

\subsection{Dynamic Program}

The dynamic program iterates over all infrastructure elements and maintains a list of time stamps on which the new train could be at that location.
However, we now have the restriction that there must exist a proceeding conflict-free path from the first station.
To be able to determine the travel time of a path, we also keep track of the departure time at the first station.
Again, all times are represented by intervals, to avoid discretizing the timescale.
More precisely, we maintain a list of pairs of intervals, where the second pair specifies the corresponding range of departure times at the start.
We require that the original departure times can be derived linearly from the first interval of possible timestamps at the current location.
Consider, for example, a pair of intervals $\left(\left(t_1^{\text{curr}}, t_2^{\text{curr}}\right), \left(t^{\text{dep}}_1, t^{\text{dep}}_2\right)\right)$.
If the train is at the current location at time $t_1^{\text{curr}} + \lambda \left(t_2^{\text{curr}} - t_1^{\text{curr}}\right)$ for some $\lambda \in [0, 1]$, we require that the corresponding departure at the start equals $t^{\text{dep}}_1 + \lambda \left( t^{\text{dep}}_2 - t^{\text{dep}}_1 \right)$, i.e., there is a linear correspondence.
If the mapping becomes more complex, the interval must be split up.
For each time stamp, we only keep track of the latest possible departure time at the start, and we ensure that all intervals corresponding to the current location are disjoint.
We denote these lists of pairs of intervals with $X$.
The following overview explains for which settings we compute those lists.
We use $P = \{\text{R}, \text{S}\}$ to denote the two stopping patterns.
\begin{itemize}
    \item For each station $s \in \mathcal S$, track $j \in f(s)$ and $p \in P$, we compute $X_{\mathcal S}[s, j, p]$.
    \item For each transition from a station $s \in \mathcal S$ with track $j \in f(s)$ to a segment $\ell \in \mathcal L$ with track $k \in f(\ell)$, and each $p \in P$, we compute $X_{\mathcal T}[s, j, \ell, k, p]$, if the segment track can be reached from the station track.
    \item For each segment $\ell \in \mathcal L$, track $k \in f(\ell)$ and $p_1, p_2 \in P$, we compute $X_{\mathcal L}[\ell, k, p_1, p_2]$.
    \item For each transition from a segment $\ell \in \mathcal L$ with track $k \in f(\ell)$ to a station $s \in \mathcal S$ with track $j \in f(s)$, and for each $p \in P$, we compute $X_{\mathcal T}[s, j, \ell, k, p]$, if the segment track can be reached from the station track.
\end{itemize}
For intermediate values like possible arrival times at stations or segments, we use $X_1$ and $X_2$.
Let $u$ be the first station to be traversed with tracks $f(u)$.
For any track $j \in f(u)$, we initialize $X_{\mathcal S}[u, j, \text{S}]$ with $F_{\mathcal S}(u, j)$ and initialize $X_{\mathcal S}[u, j, \text{R}] = \emptyset$.
The procedure uses the following subroutines that operate on pairs of interval lists, i.e., a list of possible time stamps at a location and the corresponding list of departure intervals at the origin.
\begin{itemize}
    \item The \emph{union} of lists of interval pairs returns all intervals contained in any of the lists.
    When intervals overlap, it keeps the interval with the latest departure time at the origin.
    We use this procedure, for example, after stations with multiple tracks to merge the tables for the individual tracks together.
    \item The \emph{intersection} of a list of interval pairs with lists of free spaces returns all intervals (with corresponding original departures) that are contained in all free spaces.
    The intersection is used in all steps to remove times that conflict with existing trains.
    \item The \emph{shifting} of a pair of interval lists by $d$ within segment free spaces $F$ creates two kinds of intervals.
    First, the original intervals are literally shifted by $d$, although possibly truncated to ensure that they stay in the free spaces $F$.
    However, trains can also run slower, so additional intervals are created for time stamps that could not be reached at full speed.
    Observe that these intervals have a constant original departure time.
    This operation is used when deriving the possible arrival times at the next station from those at the previous station.
    \item The \emph{extension} of a pair of interval lists within station free spaces $F$ adds all intervals on which the train could be waiting at the station, again with a constant original departure time.
\end{itemize}
With these subroutines, we can iterate the following procedure over all segments in $\mathcal L$, which is assumed to be ordered properly.
When $\ell$ is the current segment, we let $s_1$ denote the previous station and $s_2$ the next station.
We let $x$ denote the train to be inserted.

\begin{itemize}
    \item (\emph{Filtering transitions.}) For each track $j \in f(s_1)$, for each track $k \in f(\ell)$ that can be reached from track $j$ in station $s_1$, and for each $p \in P$, let $X_{\mathcal T}[s_1, j, \ell, k, p]$ be the intersection of $X_{\mathcal S}[s_1, j, p]$ and the free intervals $F_{\mathcal T}(s_1, j, \ell, k)$.
    \item (\emph{Merging obtained intervals and filtering with segment free intervals.}) For each track $k \in f(\ell)$ and $p_1, p_2 \in P$, let $X_{\mathcal L, 1}[\ell, k, p_1, p_2]$ be the union of $X_{\mathcal T}[s_1, j, \ell_i, k, p_1]$ over all tracks $j \in f(s_1)$ that can reach track $k$.
    Let $X_{\mathcal L, 2}[\ell, k, p_1, p_2]$ be the intersection of $X_{\mathcal L, 1}[\ell, k, p_1, p_2]$ with free intervals $F_{\mathcal L}(\ell, k)$ (we use the free intervals at the $s_1$-end of the segment).
    \item (\emph{Enforcing minimum running times at segments.}) Recall that $d_{x, s_1, s_2}^{p_1, p_2}$ is the minimum running time for train $x$ from station $s_1 \in \mathcal S$ to station $s_2 \in \mathcal S$, where $p_1$ and $p_2$ indicate the stopping patterns at the stations.
     Then for every track $k \in f(\ell)$ and $p_1, p_2 \in P$, we let $X_{\mathcal L}[\ell, k, p_1, p_2]$ be the shifting of $X_{\mathcal L, 2}[\ell, k, p_1, p_2]$ within $F_{\mathcal L}(\ell, k)$ by at least $d_{x, s_1, s_2}^{p_1, p_2}$.
    Lastly, $X_{\mathcal L}[\ell, k, p_2]$ merges the interval lists corresponding to stopping at $s_1$ with those corresponding to running through $s_1$.
    \item (\emph{Filtering transitions.})
    Obtain $X_{\mathcal T}[\ell, k, s_2, j, p]$ for $k \in f(\ell), j \in f(s_2)$ and $p \in P$, analogous to (1).
    \item (\emph{Merging, filtering and extending track occupations for allowed stops.})
    For $k \in f(s_2)$, let $X_{\mathcal S, 1}[s_2, j, p]$ be the union of $X_{\mathcal T}[\ell, k, s_2, j,  p]$ over all $k \in \ell$ that can reach track $j$, intersected with the free intervals $F_{\mathcal S}(s_2, j)$.
    If $p = \text{S}$, then $X_{\mathcal S, 2}[s_2, j, p]$ is the extension of $X_{\mathcal S, 1}[s_2, j, p]$ in $F_{\mathcal S}(s_2, j)$, otherwise they are equal.
    Lastly, if $X_{\mathcal S}[s_2, j, p]$ is nonempty due to an alternative route, we merge $X_{\mathcal S}[s_2, j, p]$ and $X_{\mathcal S, 2}[s_2, j, p]$ as our new $X_{\mathcal S}[s_2, j, p]$.
    Otherwise, it is equal to $X_{\mathcal S, 2}[s_2, j, p]$.
\end{itemize}

\subsection{Constructing the Corresponding Paths}
Once all tables are filled, we can read the tables backwards to find the actual paths.
Recall that these tables contain pairs of intervals: the first interval indicates when the train can reach that station, while the second one keeps track of the corresponding departure times at the first station.
Therefore, at the destination, we must find the first possible arrival time for each feasible original departure time in the list of interval pairs.
This yields the first departure time and last arrival time of all non-dominated paths.
The time at the second-last station can be obtained by scanning the corresponding tables for an interval that can both reach the last station at the determined arrival time, while departing at the start at the determined departure time.
The process can be sped up by maintaining pointers to earlier intervals.
When there is a tie between two different routes through the network, we use an arbitrary one.
In other cases of a tie, we take the earliest possible time stamp to reach each location.
This is in line with Trafikverket's policy of allowing trains to continue to new locations as much as possible.

\section{Results}\label{sec:results}
We evaluate the performance of our algorithm on a use case on the Swedish railway network.
We find the non-dominated paths for a new freight train on the 180 km long railway line from Göteborg Sävenäs to the Norwegian border at Kornsjö.
Once the yard is left, the line is double track from Göteborg Marieholm to Öxnered, and single track from Öxnered to Kornsjö.
On a double-track segment, we only use the northbound track.
We assume that all single-track segments consist of only one block, and that all stations allow simultaneous arrivals from different directions on different tracks.
For conflicting trains in a transition, only trains using the same track combination were considered. 
More complicated combinations can be added manually to avoid other conflicts, or to avoid simultaneous arrivals if necessary.
We consider the time window from 07:00 until 14:00 on January 20th, 2021.
All source code is available on \url{https://github.com/david-dekker-liu/path-insertion}.

\begin{figure}
\centering
\includegraphics[width=0.9\textwidth]{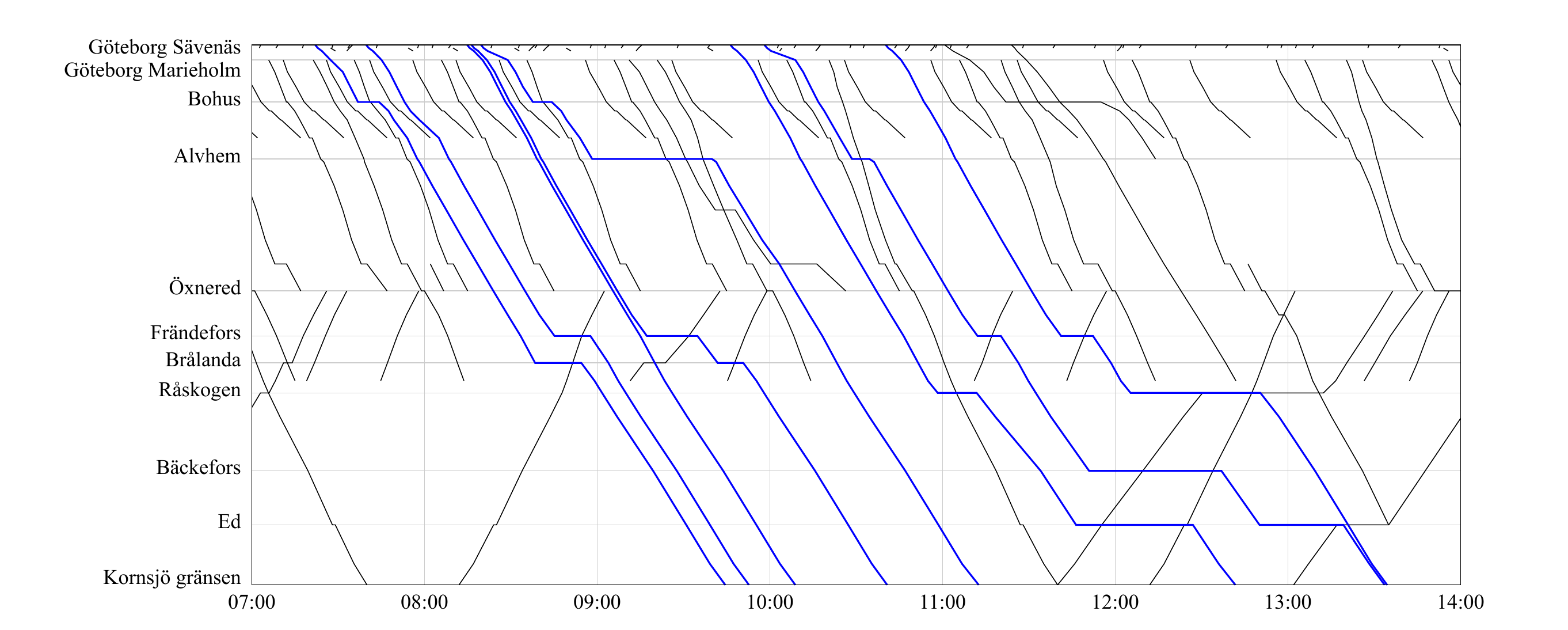}
\caption{The eight candidate paths for a freight train from Göteborg Sävenäs to Kornsjö. Locations where trains can overtake or meet each other are indicated with horizontal, grey lines. One existing path uses side tracks for which the opposing tracks have to be crossed. We do not consider this behavior, as crossing opposing tracks consumes relatively much capacity, but it can easily be included in the algorithm.}\label{fig:paths}
\end{figure}

Figure~\ref{fig:paths} displays the obtained timetable.
Black train paths were already in the timetable, and the blue paths are candidates for the freight train.
On double-track segments between Göteborg Marieholm and Öxnered, only trains on the northbound track are displayed.
Since the number of paths was fairly limited, the candidate paths were derived from the dynamic programming tables manually.

The algorithm took $0.3$ seconds on average over five runs to compute the paths.
The preprocessing steps were not optimized and took $10.4$ seconds on average, but we want to emphasize that these computations are not necessary if the free intervals are maintained with the timetable.
Observe that, by construction of the algorithm, the paths typically have the same robustness: if a train would wait longer than necessary after a stop, the resulting path would be dominated by the path that departs first.
Only in case of a range of parallel paths that do not stop, one could pick a path with a larger robustness.
In Figure~\ref{fig:paths}, the third path is such an example, even though these parallel paths are within forty seconds of each other.
Notice that it is likely that this path is used on other days, as an opposing train seems to wait without clear reason.

Figure~\ref{fig:paths} also illustrates the advantages of obtaining all non-dominated paths, rather than only the optimal one. If a railway undertaking would like to arrive close to 13:30 at the Norwegian border, the seventh path might be considered as optimal.
However, the eigth (and last) path in the figure arrives less than one minute later, while being over forty minutes faster.
Similarly, the third, fourth and fifth option have very similar departures in Göteborg, but differ more than an hour in Kornsjö.

\begin{figure}[ht!]
\centering
\begin{subfigure}{.45\textwidth}
  \centering
  \includegraphics[width=\linewidth]{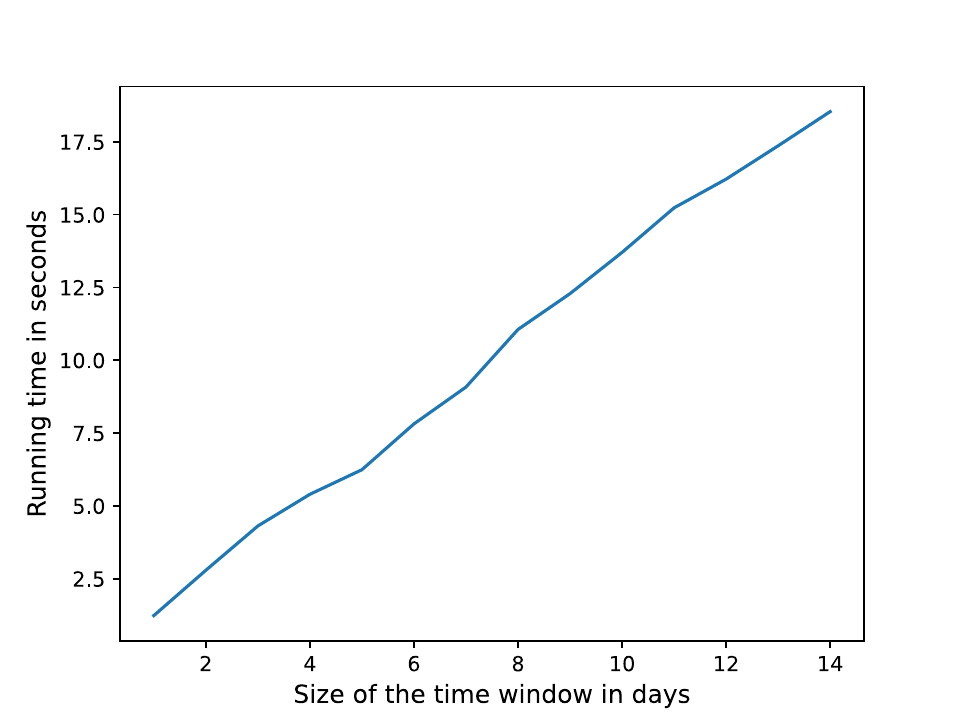}
  \caption{Running time of the path insertion algorithm for larger time windows, excluding preprocessing. All time windows start on 00:00 at January 18th, 2021, and span an integer number of days. The weekend days are the 6th, 7th, 13th and 14th day.}
  \label{fig:time-windows}
\end{subfigure}%
\hfill
\begin{subfigure}{.45\textwidth}
  \centering
  \includegraphics[width=\linewidth]{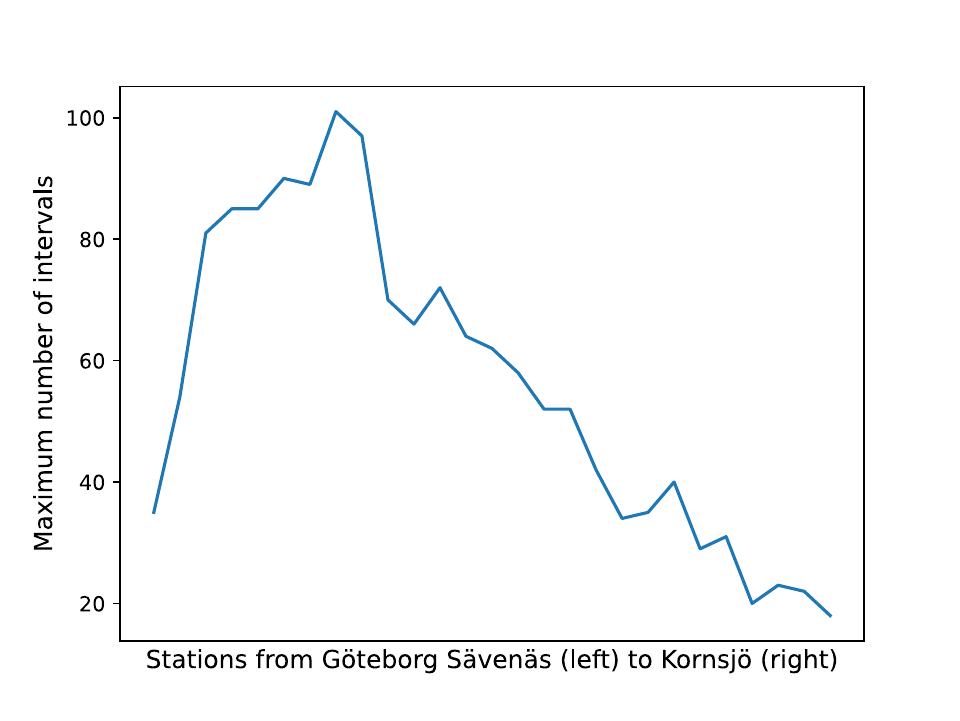}
  \caption{Size of the dynamic programming tables for the original time window of seven hours. On each location, we sum the size of the tables corresponding to stopping and running through.
  If a location has multiple usable tracks, we only display the largest interval count over the possible tracks.
  }\label{fig:table-size}
\end{subfigure}
\caption{Evaluation of the path insertion algorithm.}\label{fig:alg-evaluation}
\end{figure}

Figure~\ref{fig:alg-evaluation} illustrates the running time and storage properties of the algorithm.
We used time windows of several days to analyze the running time, and averaged five runs per window size again.
While that is not a realistic use case, the results of smaller time windows might be influenced by the varying number of trains during the day.
As expected, Figure~\ref{fig:time-windows} suggests that the dependency on the size of the time window is linear, while the table sizes in Figure~\ref{fig:table-size} seem unrelated to the network size.
The decrease in the number of intervals, however, is interesting.
We expect that these are caused by longer stretches without side tracks on double-track lines or meeting stations on single-track lines.
The difference between double and single track is not the only cause: by the time we reach the single-track line in Öxnered, the number of intervals is already down to 52 from 101 at its peak.
It is reassuring that the number of intervals never increases significantly after the start.
Therefore, we expect that the running time depends linearly on the network size in the worst case, as the number of intervals at a station roughly corresponds to the number of operations that has to be executed to transverse a segment leaving the station.

\section{Conclusions and Future Work}
We presented a fast and scalable dynamic program for adding a train path to a timetable in a railway network.
We tested its performance on the railway line from Göteborg Sävenäs to the Norwegian border in Kornsjö, a line with both double-track and single-track segments.
After preprocessing, the algorithm is indeed fast, and the size of the computed tables stays small.
Some interesting follow-up questions would focus on extensions to multiple trains, or on extensions to microscopic infrastructure details.

\section*{Acknowledgements}
We would like to thank Trafikverket, and Kristian Persson in particular, for the good cooperation.
The project was funded by Trafikverket (Dnr. TRV 2020/72700).







\bibliographystyle{elsarticle-harv.bst}
\bibliography{TRPRO_EWGT_2024}










\clearpage

\end{document}